\documentclass[aps,prl,superscriptaddress,letterpaper,floats,floatfix,letterpaper,floatfix]{revtex4-1}

\usepackage{amsmath}
\usepackage[pdftex]{graphicx}
\usepackage[thinspace]{SIunits}
\usepackage{epsfig}
\usepackage[usenames]{color}

\newcommand{\nn}{\nonumber}

\newcommand{\bs}{\mathbf{s}}

\newcommand{\dg}{\dagger}

\newcommand{\be}{\begin{eqnarray}}
\newcommand{\ee}{\end{eqnarray}}
\newcommand{\la}{\langle}
\newcommand{\ra}{\rangle}
\newcommand{\rar}{\rightarrow}
\newcommand{\da}{\downarrow}
\newcommand{\ua}{\uparrow}

\begin{document} 
 
\title{ Dynamic Symmetries and Quantum Nonadiabatic Transitions }

\author{Fuxiang Li}
 \affiliation{Center for Nonlinear Studies, Los Alamos National Laboratory,  Los Alamos, New Mexico 87545 USA}
 \affiliation{Theoretical Division, Los Alamos National Laboratory, Los Alamos, New Mexico 87545, USA}
\author{Nikolai A. Sinitsyn}
 \affiliation{Theoretical Division, Los Alamos National Laboratory, Los Alamos, New Mexico 87545, USA}
 
\date{\today}

\begin{abstract}
Kramers degeneracy theorem is one of the basic results in  quantum mechanics. According to it, the time-reversal symmetry makes each energy level of a half-integer spin system at least doubly degenerate, meaning the absence of transitions or scatterings between degenerate states if the Hamiltonian does not depend on time explicitly. We generalize this result to the case of explicitly time-dependent spin Hamiltonians. We prove that for a spin system with the  total spin being a half integer, if its Hamiltonian and the evolution time interval are symmetric under a specifically defined time reversal operation, the scattering amplitude between an arbitrary initial state and its  time reversed counterpart is exactly zero. We also discuss applications of this result to the multistate Landau-Zener (LZ)  theory.
\end{abstract}

\maketitle
\section{Introduction}

Adiabatic approximation is widely used in nonstationary quantum mechanics. According to it, a system's state vector changes with time so that it remains an eigenstate of an effective slowly time-dependent Hamiltonian. This approximation breaks down near moments of time when different instantaneous discrete energy levels of a quantum system approach too close to  each other. Such events lead to quantum nonadiabatic transitions, which considerably complicate system's dynamics. In \cite{chernyak-prl}, Chernyak {\it et al} developed the theory of  nonadiabatic transitions for a  wave packet passage through a Dirac cone intersection of two energy surfaces. Such an intersection is also known in molecular physics as a diabolical point.  Originally, the work  \cite{chernyak-prl} was motivated by applications to molecular physics, however, recent discoveries of novel Dirac materials, such as graphene, topological insulators, Weyl semimetals, and transition metal  dichalcogenides, have triggered numerous conceptually similar studies of wavepacket scattering from a Dirac point in solid state systems \cite{dirac-lz1,dirac-lz2,sinitsyn2015solvable}. 

The simplicity of the theory in  \cite{chernyak-prl} follows from the fact that, in the ballistic limit, the effective Hamiltonian of the problem can be reduced to the widely known Landau-Zener (LZ) model for a spin-1/2 coupled to a linearly time-dependent magnetic field with the Hamiltonian
\be
\hat{H}(t) = \beta t \hat{\sigma}_z + g \hat{\sigma}_x,
\label{lz1}
\ee 
where $\hat{\sigma}_z$ and   $\hat{\sigma}_x$ are 2$\times$2 Pauli matrices. Corresponding Schr\"odinger equation is exactly solvable in terms of parabolic cylinder functions, whose asymptotic behavior are known. For example, if the evolution starts in the $\uparrow $ spin state at $t\rightarrow -\infty$ then the probability to  find this spin flipped at time $t\rightarrow +\infty$, after the evolution with the Hamiltonian (\ref{lz1}), is given by a simple LZ formula:
\be
P_{\downarrow  \uparrow}  = 1-e^{-\pi g^2/\beta}. 
\label{lz2}
\ee

Application of the theory  \cite{chernyak-prl} to conduction electrons, however, is  restricted because  electrons in Dirac materials also experience considerable spin-orbit coupling. In addition to electron-hole index, true electronic spin becomes coupled to degrees of freedom of the Dirac Hamiltonian and, hence, should be taken into account. Also, Dirac cones appear as degenerate pairs, which  can be coupled by potentials of  short range impurities which make them nonseparable \cite{sinitsyn2015solvable}  in some situations. 
Therefore, even after employing ballistic approximation, which was used in \cite{chernyak-prl}, the effective Hamiltonian is generally represented not by a 2$\times$2 but by a 4$\times$4, 8$\times$8 or even higher dimensional matrix, 
which can be mapped to more complex than (\ref{lz1}) interacting spin models with explicitly time-dependent parameters \cite{sinitsyn2015solvable}.  Corresponding Shr\"odinger equation  would be  equivalent to a higher than 2nd order differential equation with time-dependent coefficients.
Connections between asymptotics  of solutions of such equations are generally unknown.

 Dirac materials are only few of many examples that motivate developments of the theory of {\it multistate} nonadiabatic transitions. One of the main questions in this theory can be formulated as the following quite a general scattering problem: Given that $\hat{H}(t)$ is the N$\times$N matrix Hamiltonian with time-dependent parameters such that there is a time-interval $(-T,T)$, beyond which nonadiabatic transitions are suppressed (e.g., because  time-dependence of parameters is restricted to this time interval), then what are the probabilities of transitions between stationary states that can be defined for times $t<-T$  and stationary states at $t>T$?

Apart from Dirac materials, such scattering problems are frequently found in true spin systems that are driven by time-dependent magnetic fields. For example, molecular nanomagnets are molecules with a number of magnetic ions, such as Fe, Mn, Co, whose spins behave as a single collective spin ${\bf S}$ (typically larger than 1/2) at sufficiently low temperature. For $S>1/2$, the crystal anisotropy effect and action of a time-dependent field, ${\bf B}(t)$, is described by the Hamiltonian
\be
\hat{H}={\bf B}(t) \cdot {\bf \hat{S}}- a\hat{S}_z^2 + b\hat{S}_x^2+c\left[(\hat{S}^+)^4 + (\hat{S}^-)^4\right]+\ldots, 
\label{nanom1}
\ee
where $\hat{S}_{\alpha}$ are operators of the collective spin, parameters $a$ and $b$ characterize the leading order quadratic anisotropies, and other terms describe higher order corrections to the anisotropy that are allowed by the time-reversal symmetry at ${\bf B}=0$. Note that we absorbed the Lande g-factor and Bohr magneton into the definition of the magnetic field. 
Magnetization hysteresis curves that  were obtained  by application of a periodic in time high amplitude magnetic field to a nanomagnet array
showed multiple steps \cite{Wernsdorfer1999, Vijayaraghavan2009} that were explained as due to successive LZ transitions at avoided crossings of
different pairs of energy levels.

Another physical example is the spin of electron in a quantum dot or the  electronic states of an NV-center in diamond interacting with a nuclear spin \cite{NV-LZ}. 
A localized electron state can be controlled either by magnetic or by optically induced  field pulses. However, coupling of the electronic  system to nuclear spin(s) complicates the dynamics. For example, if an electronic spin qubit in a quantum dot is coupled to a nuclear spin, the Hamiltonian can be written as
\be
\hat{H}= {\bf B}(t) \cdot {\bf \hat{\sigma}} +  g \mathbf{\hat{\sigma}} \cdot {\bf \hat{S}} + \cdots,
\label{NV}
\ee
where $\hat{\sigma}_{\alpha}$ are Pauli matrices acting in the qubit subspace and $\hat{\bf S}$ is the nuclear spin operator. Not shown terms, which are allowed by the time-reversal symmetry, describe hyperfine coupling anisotropy and quadrupole coupling of the nuclear spin to electric fields produced by local strains. One can also include weak coupling of the magnetic field to nuclear spins.

Behavior of quantum systems with explicitly time-dependent Hamiltonians, such as (\ref{nanom1}) and (\ref{NV}), is usually hard to explore because of multiple interfering nonadiabatic transitions and the lack of standard conservation laws in nonstationary quantum mechanics. 
In this article we show, however, that some of the ideas used in stationary quantum mechanics can be extended to  quantum models with explicitly time-dependent parameters. 

It has been long known that in half-integer spin systems with the time-reversal symmetry  all states are  Kramers degenerate, and tunneling between degenerate states is forbidden \cite{Delft92, Garg1995}. We  will show that for some explicitly time-dependent situations, half-integer multi-spin systems preserve this property in the sense that transitions between Kramers doublet states have identically zero probabilities after completion  of the time-dependent control field pulses.


\section{The Model} 

The most general model that we will consider describes  an explicitly time-dependent system with $N$ interacting spins, and with a Hamiltonian that can be written in the following form:
\be
\hat{H}(t) = \sum_{n} \sum_{\alpha_m=x, y, z} \sum_{ j_m=1, \ldots, N} C^{j_1,\ldots, j_n}_{\alpha_1, \ldots, \alpha_n}(t) \hat{S}^{j_1}_{\alpha_1} \hat{S}^{j_2}_{\alpha_2} \cdots \hat{S}^{j_n}_{\alpha_n}, 
\label{eq:general_H}
\ee
where $\hat{S}^{j_m}_{\alpha_m}$ is the projection operator of $j_m$th spin  on $\alpha_{m}$th-axis.  The total spin of the system is assumed to be half-integer.
 If a spin is larger than 1/2, it is possible to have identical indexes,  $j_m=j_{m'}$ and $\alpha_{m} = \alpha_{m'}$ for $m \ne m'$, in the coefficients $C^{j_1,\ldots, j_n}_{\alpha_1, \ldots, \alpha_n}(t)$ in (\ref{eq:general_H}), which  allows higher powers of the same spin operator to be included.
For each term in (\ref{eq:general_H}), $\hat{S}^{j_1}_{\alpha_1} \hat{S}^{j_2}_{\alpha_2} \cdots \hat{S}^{j_n}_{\alpha_n}$ is the product of $n$ spin operators, where $n$ runs to mark all possible independent combinations of such spin operator products. 

We further restrict time dependent coefficients $C^{j_1,\ldots, j_n}_{\alpha_1,  \ldots, \alpha_n}(t) $ so that they satisfy the following symmetry:
\be
&&C_{\alpha_1, \cdots, \alpha_n}^{j_1,\ldots, j_n}(-t)  = - C_{\alpha_1, \ldots, \alpha_n}(t) , ~~ {\rm if}~ n ~ {\rm is ~odd},  \nn \\
&&C_{\alpha_1,  \cdots, \alpha_n}^{j_1,\ldots, j_n}(-t) =  C_{\alpha_1, \ldots, \alpha_n}(t) , ~~ {\rm if}~ n ~ {\rm is ~ even},  \label{eq:coeff}
\ee 
and assume that evolution proceeds during time interval, $t\in(-T,T)$, which is symmetric under reflection from $t=0$. 

 For a spin system, one can always define a time-reversal operator \cite{Sakurai}
\be
\Theta = \exp(- i\pi \hat{S}_y) \hat{K}, \label{eq:theta}
\ee
where $\hat{S}_{y}$ is the $y$ component of the total spin operator: $\hat{S}_y = \sum_{i=1}^{N}  \hat{S}^{i}_{y}$, where $N$ is the total number of spins in the system; $\hat{K}$ is the  operator that forms the complex conjugate of any coefficient that multiplies a ket-vector on which $\hat{K}$ acts. Therefore, $\Theta$ is an antiunitary operator, which means that for two given states $|\alpha \ra$ and $|\beta \ra$, $\la \Theta \alpha | \Theta \beta \ra = \la \beta | \alpha \ra$. Under action of $\Theta$, each spin operator changes its sign, i.e. $\Theta \hat{S}^i_{\alpha_i} \Theta^{-1} = - \hat{S}^i_{\alpha_i}$ (Appendix).  
Therefore, although the Hamiltonian (\ref{eq:general_H}) is explicitly time-dependent, it satisfies the following symmetry:
\be
\Theta \hat{H}(t) \Theta^{-1} = \hat{H}(-t).  \label{eq:sym_H}
\ee

It is well known that, for half-integer spin systems, $\Theta^2=-1$, while for integer spin systems, $\Theta^2=1$. This property has been used to prove the Kramers degeneracy theorem \cite{Sakurai}. 
Now, we consider the explicitly time-dependent situation (\ref{eq:coeff}), which is described by nonstationary Schr\"odinger equation $id\Psi(t)/dt = \hat{H}(t) \Psi(t)$, where $\Psi$ is the state vector. By analogy with the proof of the Kramers degeneracy theorem, we can prove the following its extension.


\section{No-Scattering Theorem}
The {\it  No-Scattering Theorem} states that  for a  spin system with the time-dependent Hamiltonian $\hat{H}(t)$ satisfying the symmetry (\ref{eq:sym_H}), if the total spin of the system is half-integer, then the scattering amplitude from an initial state $|\Psi(-T) \ra$ to its time-reversed counterpart at the end of the evolution, i.e., to  $|\Psi' \ra=\Theta | \Psi(-T) \ra$, is exactly zero: 
\be
S \equiv \la \Theta \Psi(-T) | \hat{U}(T, -T) | \Psi(-T)  \ra =0, \label{eq:S}
\ee
where $\hat{U}(T, -T)$ is the time evolution operator  from time $-T$ to time $T$ with the Hamiltonian $\hat{H}(t)$. 

 {\underline{Proof}}:  First, let us explore behavior of  the evolution operator $\hat{U}$ under the time reversal operation. The standard expression for the evolution operator can be written as $\hat{U}(T ,-T) = { \cal{T}} e^{-i\int_{-T}^T H(\tau) d\tau}$ where ${\cal{T}}$ is the time ordering operator. For our purpose, we divide the time from $-T$ to $T$ into $2n$ time steps, with time points denoted as: $t_j = j T/n$ where $j$ runs from $-n+1/2$ to $n-1/2$ in unit steps. In the limit of $n \rar \infty $, the time evolution operator can be expressed as the products of evolution operators for each small time step of  size $dt=T/n$:
\be
\hat{U}(T, -T) = e^{-i \hat{H}(t_{n}) dt} \cdots e^{-i \hat{H}(t_j)dt} \cdots e^{-i \hat{H}(t_{-j})dt} \cdots e^{-i \hat{H}(t_{-n}) dt} .  \label{eq:u}
\ee
Let us recall that $t_{-j} = - t_j$. Under the time reversal operation, for each term in (\ref{eq:u}), we have then $\Theta e^{-i \hat{H}(t_j)dt} \Theta^{-1} = e^{i \hat{H}(t_{-j})dt} $. Therefore, after inserting $\hat{1} \equiv \Theta \Theta^{-1}$ between each pair of exponents in (\ref{eq:u}) we find
\be
\Theta \hat{U}(T, -T) \Theta^{-1} = e^{i \hat{H}(t_{-n}) dt}   \cdots e^{i \hat{H}(t_{-j})dt} \cdots  e^{i \hat{H}(t_{j})dt}  \cdots e^{i \hat{H}(t_{n}) dt},  \label{eq:ud}
\ee
which is the same as the hermitian conjugate of  (\ref{eq:u}), i.e.,
\be
\Theta \hat{U}(T, -T) \Theta^{-1} = \hat{U}^{\dg}(T, -T). \label{eq:Usym}  
\ee
 
Let us denote, for simplicity, $\Psi \equiv \Psi(-T)$, and $\hat{U}\equiv \hat{U}(T, -T)$. We insert two pairs of $\Theta^{-1} \Theta$ in the expression of $S$ given in (\ref{eq:S}):
\be
S = \la \Theta \Psi | \Theta^{-1} \Theta \hat{U} \Theta^{-1} \Theta \Psi \ra . 
\ee
For a half-integer total spin system,  we have $\Theta^2=-1$, i.e. $\Theta^{-1} = - \Theta$. Substituting the first $\Theta^{-1}$ with $-\Theta$ and using the property (\ref{eq:Usym}), we find
\be
S = - \la \Theta \Psi | \Theta (\Theta \hat{U} \Theta^{-1}) \Theta \Psi \ra = - \la \Theta \Psi | \Theta \hat{U}^{\dg} \Theta \Psi \ra .
\ee
Since $\Theta$ is an anti-unitary operator, for any two states $|\alpha \ra$ and $| \beta \ra$,  one has $\la \Theta \alpha | \Theta\beta \ra = \la \beta | \alpha \ra$. Treating $\la \alpha | \equiv \la \Psi |$ and $| \beta \ra \equiv | U^{\dg} \Theta \Psi \ra$, we obtain
\be
S=- \la\hat{U}^{\dg}   \Theta \Psi | \Psi \ra = - \la  \Theta \Psi | \hat{U} \Psi \ra = -S ,
\ee
which can be resolved only by seting $S=0$, which proves the theorem.

\section{Applications: Multistate Landau-Zener theory} 

Some of the most frequently encountered realizations of the spin Hamiltonian (\ref{nanom1})-(\ref{NV}) correspond to the linear sweep of the magnetic field, $B=\beta t$, where $t$ is time and $\beta$ is a constant sweeping rate. At large negative or positive times, energy levels of such a spin system are split so that transitions between different spin projections on the magnetic field axis do not happen. In the matrix form, such Hamiltonians belong to the class of, so called, multistate LZ model \cite{brundobler1993s} that describes time-dependent evolution with  the Hamiltonian of the form 
\be
\hat{H}(t) = \hat{A} + \hat{R} t ,
\label{mulLZ2}
\ee
where $\hat{A}$ and $\hat{R}$ are time-independent Hermitian $N \times N $ matrices. Eigenstates of the matrix $\hat{R}$ are called diabatic states. They are almost the eigenstates of the full Hamiltonian at $t\rightarrow \pm \infty$.
The goal of the multistate LZ theory is to find the scattering $N \times N$ matrix $\hat{S}$, whose element $S_{nn'}$ is the amplitude of the diabatic state $n$ at $t \rar + \infty$, given that at $t \rar -\infty$ the system was in the $n'$-th diabatic state. From the scattering matrix $\hat{S}$  one can obtain transition probabilities $P_{n\rightarrow n'} \equiv |S_{n'n}|^2$ between pairs of diabatic states.  

Generally, for $N > 2$, the analytical solution of the Schr\"odinger equation with the Hamiltonian (\ref{mulLZ2}) is unknown. 
Nevertheless, a number of exactly solvable cases with specific forms of matrices $\hat{A}$ and $\hat{R}$ have been derived \cite{demkov1995, sinitsyn2015solvable, sinitsyn2015exact,sinitsyn2004counterintuitive, demkov1995}.  
In this section, we apply the No-Scattering Theorem to illustrate some of the properties of multistate LZ spin models.

\subsection{Spin-3/2 with quadrupole coupling and time-dependent magnetic field}
Consider a nanomagnet with spin $S=3/2$ in  a linearly changing magnetic field and a static anisotropy field. The Hamiltonian can be most generally written as: 
\be
\hat{H}_1= g_1(\hat{n}\cdot { \hat{\bf S} })^2 + h_1 t \hat{S}_z  + h_2 t \hat{S}_z^3 + g_2 (\hat{n}_{\perp}\cdot \hat{\bf S})^2. \label{H3_2}
\ee
Here  $\hat{\bf S} \equiv  (\hat{S}_x, \hat{S}_y, \hat{S}_z)$ are the  spin-$3/2$ operators, and we switched to the diabatic basics, in which the magnetic field is coupled only to $\hat{S}_z$.  Unit vectors $\hat{n}$ and  $\hat{n}_{\perp}$,  such that $\hat{n}_{\perp} \cdot \hat{n} = 0$, describe relative directions of the  anisotropy axes and the direction of the magnetic field.  Constant parameters $g_1$ and $g_2$  describe quadratic anisotropy couplings and $h_1$, $h_2$ are the sweeping rates describing coupling to the linearly time-dependent external field. One can expect that $h_{2}=0$ but it can become nonzero as a result of some renormalization of parameters, similarly to the renormalization of the Lande g-factor.

One can parametrize $\hat{n} = (\cos\phi, 0, \sin\phi)$, and $\hat{n}_{\perp} = (-\sin\phi \sin\theta, \cos\theta, \cos\phi\sin\theta)$ with $\theta$ and $\phi$ two angles.  For simplicity, we will consider the special case $\theta=0$.
The time reversal operator is $\Theta = e^{-i \pi \hat{S}_y} \hat{K}$. Since  $\Theta \hat{H}_1(t) \Theta^{-1} = \hat{H}_1 (-t)$, the identity (\ref{eq:S}) applies for this system. 
In the diabatic basis $ \left(| \frac{3}{2} \ra,  | -\frac{3}{2} \ra, | \frac{1}{2} \ra, | - \frac{1}{2}\ra \right)$ of eigenvectors of the spin operator $\hat{S}_z$, the Hamiltonian is a 4$\times $4 matrix: 
\be
\hat{H}_1=\begin{pmatrix}
  \beta_1 t + \Delta_1  &0  & \gamma_2 & \gamma_1  \\
  0 &-\beta_1 t + \Delta_1  & \gamma_1 & -\gamma_2 \\
    \gamma_2 & \gamma_1 & \beta_2 t + \Delta_2  & 0    \\
  \gamma_1 & - \gamma_2  & 0&  -\beta_2 t + \Delta_2   \\  
\end{pmatrix} 
\label{ham32},
\ee 
with $\beta_{1,2}$ and $\Delta_{1,2,3}$  determined by  parameters $h_{1,2}$, $g_{1,2}$, and $\phi$: 
$\beta_1=\frac{3}{2} h_1  + \frac{27}{8} h_2 , \,  \beta_2 = \frac{1}{2} h_1  + \frac{1}{8}h_2,  \,
\Delta_1 = \frac{3}{2}(g_1 + g_2) - \frac{3}{4} (g_1 -g_2) \cos(2\phi),  \,
 \Delta_2 = g_1 +  g_2 + \frac{3}{4} (g_1 -g_2) \cos(2\phi),  \,   \gamma_2=\frac{\sqrt{3}}{2}(g_1-g_2)\sin(2\phi), \gamma_1=\frac{\sqrt{3}}{4} [ g_1+g_2 + (g_1 -g_2) \cos(2\phi)]$. 

The No-Scattering Theorem predicts that the scattering amplitudes between  diabatic states $|3/2\ra $ and $|-3/2\ra $, as well as between states $|1/2\ra$ and $|-1/2\ra$, are identically zero.
We can verify this prediction by noticing that the four-state LZ model  (\ref{ham32}) has already been fully solved in  \cite{sinitsyn2015solvable} for evolution from $t\rightarrow -\infty$ to $t\rightarrow +\infty$, although relation of the Hamiltonian (\ref{ham32}) to the spin-3/2 dynamics was not discussed then.  The known exact solution confirms the validity of the No-Scattering Theorem in this particular case.



\begin{figure}
\centering
\includegraphics[width=0.7\columnwidth]{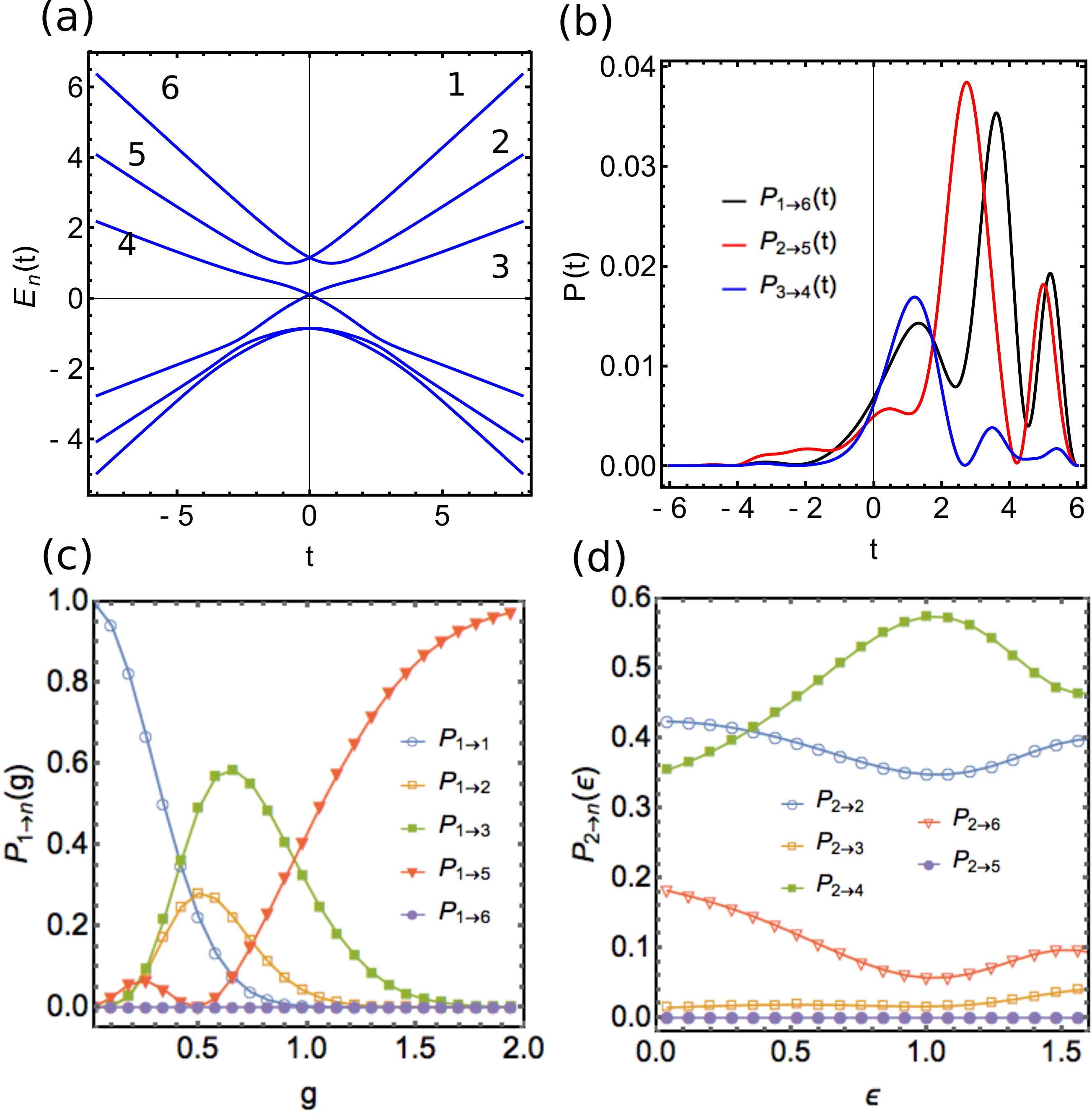}
\caption{\label{fig1} (Color Online) (a) Eigenenergy levels $E_n(t)$ as  functions of time $t$ obtained by diagonalization of the Hamiltonian (\ref{H2}).   Parameters: $\epsilon_1=1, \, \epsilon_2=0.2, \, g_1=g_2=g_4=1, \, g_3=g_5=g_6=g_7=g_8=0,\,  h_1=1.0,\, h_2=0.2,\,  h_3=0$.  (b)  Transition probabilities  from one diabatic state at time $-t_0$ to its time reversal counterpart state at time $t$ as functions of $t$. Here $t_0=6.0$ and other parameters are the same as  in (a). (c) Transition probabilities from the initial state $1$ as functions of parameter $g$ defined so that $g_1= g_2= g_4=g, \, g_3=0$. Other parameters: $\epsilon_1=1 , \, \epsilon_2=0.2,  \, h_1=1.0,\,  h_2=0.2,\,  h_3=0,\,  g_5=g_6=g_7=g_8=0$.  (d) Transition probabilities from the initial state $2$ as functions of parameter $\epsilon$ defined so that $\epsilon_1=1.0 \epsilon, \,  \epsilon_2=0.2 \epsilon$.  Other parameters: $  g_1=g_2=g_4=1., g_3=g_5=g_6=g_7=g_8=0,\,  h_1=1.0,\,  h_2=0.2, \, h_3=0$. Numerical simulations of evolution in (c)-(d) are performed for time interval $(-T,T)$, where $T=1000$. Solid lines in (c-d) are the guides for eyes (not analytical predictions).}
\end{figure}
\subsection{Spin-1/2 coupled to spin-1 in a time-dependent magentic field}
The next in complexity model of the type (\ref{NV}) is a spin-1/2 interacting with a spin-1. 
 The most general Hamiltonian in which time-independent terms conserve the time-reversal symmetry, and  in which magnetic field couplings are linearly time-dependent can be written in the diabatic basis as
\be
\hat{H}_2 &=& \epsilon_1 \hat{s}_z \hat{S}_z + \epsilon_2 \hat{S}_z^2 + g_1 \hat{s}_x \hat{S}_z + g_2 \hat{s}_x \hat{S}_x + g_3 \hat{s}_y \hat{S}_y + g_4 \hat{s}_y \hat{S}_x  +  g_5 \hat{s}_z \hat{S}_x + g_6 \hat{s}_z \hat{S}_y + g_7 \hat{s}_x \hat{S}_y + g_8 \hat{s}_y \hat{S}_z \\ \nn
&&+ h_1 t \hat{s}_z + h_2 t \hat{S}_z + h_3 t \hat{s}_z \hat{S}_z^2, 
 \label{H2}
\ee
where $\hat{s}_{x, y, z}$ and $\hat{S}_{x, y, z}$ are spin operators for, respectively, spin $1/2$ and $1$.  Parameters $\epsilon_{1}$ and $g_{i}$, $i =1, \ldots, 8$, describe interactions between  spins (e.g., anisotropic hyperfine coupling between electronic spin states and a nuclear spin), $\epsilon_2$ is the quadrupole coupling of the spin-1, and  $h_{1, 2, 3}$ describe couplings of those spins to the external linearly time-dependent magnetic or optically induced field. 

In the basis $ (1, 2, 3, 4, 5, 6) \equiv$ ($| \ua, 1 \ra$, $|\ua, 0 \ra$, $|\ua, -1\ra$, $| \da, 1\ra$, $|\da, 0 \ra$, $| \da, -1\ra$) of eigenstates of the operator $\hat{s}_z \hat{S}_z$ the Hamiltonian (\ref{H2}) has the following matrix form:
\be
\hat{H}_2 = 
\begin{pmatrix}
  \beta_1 t + \Delta_1   & \gamma_4 & 0 & \gamma_1 & \gamma_2 & 0 \\
   \gamma_4^* & \beta_2 t   & \gamma_4 & \gamma_3 & 0 & \gamma_2  \\
  0 & \gamma_4^*  & \beta_3 t -\Delta_2 & 0 & \gamma_3 & -\gamma_1 \\  
  \gamma_1 & \gamma_3^* & 0 & -\beta_3 t - \Delta_2 & -\gamma_4 & 0 \\
  \gamma_2^* & 0 & \gamma_3^*  & -\gamma_4^* & -\beta_2 t & -\gamma_4 \\
  0 & \gamma_2^*  & -\gamma_1 & 0 & -\gamma_4^* & -\beta_1 t + \Delta_1
\end{pmatrix},
\label{H22}
 \ee
where $\Delta_{1, 2} = \pm \frac{\epsilon_1}{2} + \epsilon_2$, $\beta_{1,3 }= \frac{h_1}{2} \pm h_2+ \frac{h_3}{2}$, $\beta_2=\frac{h_1}{2}$, $\gamma_1 =\frac{ g_1-g_8}{2}$ , $\gamma_{3, 2} = \frac{g_2}{2 \sqrt{2}}  \pm  \frac{g_3}{2 \sqrt{2}} - i \frac{g_4}{2\sqrt{2}} \pm i \frac{g_7}{2\sqrt{2}}$, and $\gamma_4= \frac{g_5- i g_6}{2\sqrt{2}}$. 

Here we note that for a special case with $h_2=h_3=0$ and $g_5 = g_6= 0$, one has $\beta_1=\beta_2 = \beta_3$ and $\gamma_4=0$, and then this model reduces to an exactly solvable six-state model discussed in \cite{sinitsyn2015exact}, for which predictions of the No-Scattering Theorem can be verified by comparison with the known full exact solution. However, for other values of parameters, the Hamiltonian (\ref{H22}) does not reduce to the one in \cite{sinitsyn2015exact}. Moreover, generally, the Hamiltonian (\ref{H22}) does not satisfy LZ-integrability conditions, discussed in \cite{sinitsyn2015exact}, so in order to verify predictions of the No-Scattering Theorem we  resort to numerical simulations.

In Fig.~\ref{fig1}(a), we plot the eigenenergy levels for the  model (\ref{H22}) as functions of $t$ at some arbitrary choice of constant parameters. There are three points with exact level crossings at time $t=0$, which are guaranteed by the Kramers theorem. In this case, the time reversal counterparts of states $|1\ra$, $|2\ra$ and $| 3\ra$ are, respectively, $|6\ra$, $|5\ra$ and $| 4\ra$.  In Fig.~\ref{fig1}(b), numerically calculated  transition probabilities between such pairs of levels are shown as functions of time for evolution during a finite time-interval $ (-t_0,t)$. These transition probabilities are generally nonzero and can show very complex behavior. However,  at $t=t_0$, all of them are found to be exactly zero. This illustrates the fact that zero values of transition probabilities that are guaranteed by the No-Scattering Theorem are generally achieved  only at the completion moment of the full evolution but not at intermediate time moments. 
Finally, in Figs.~\ref{fig1}(c-d), numerically calculated transition probabilities for evolution from $-\infty$ to $\infty$ (meaning that $t_0$ is chosen so large that numerically obtained transition probabilities almost do not show visible
 changes at larger $t_0$) are shown for different initial states and different choices of constant parameters. 
They confirm prediction of the No-Scattering Theorem that $P_{1\rar6}=0$ and $P_{2\rar 5} =0$ at all considered parameter choices.

\begin{figure}
\centering
\includegraphics[width=0.7\columnwidth]{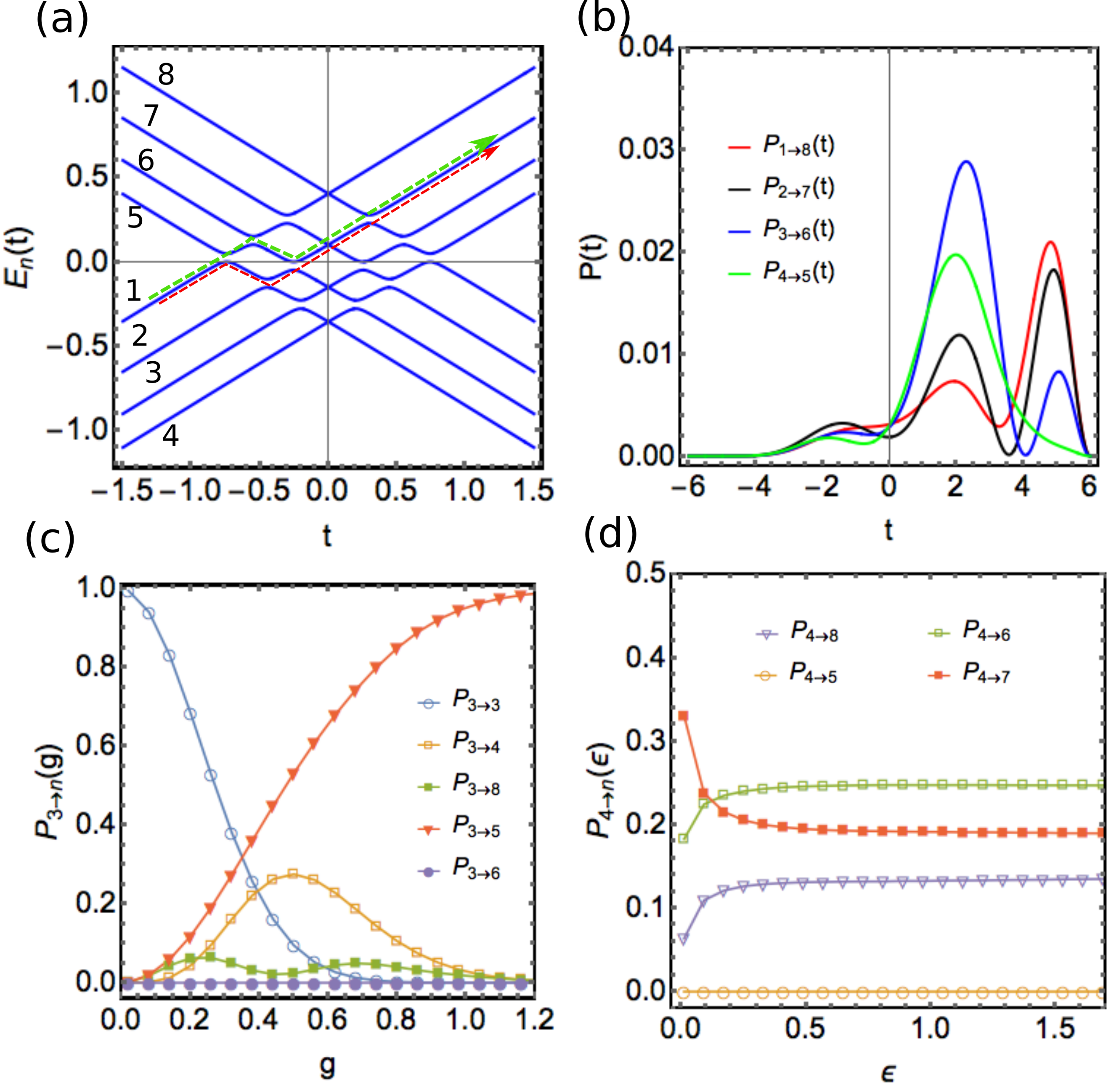}
\caption{\label{fig2} (Color Online) (a) Eigenenergy levels $E_n(t)$ as  functions of time $t$ obtained by diagonalization of the Hamiltonian (\ref{H3}).  Red and green dashed lines denote  two semiclassical trajectories connecting the state $1$ at $t \rar -\infty$ to the state $2$ at $t \rar + \infty$.  Parameters: $h_1 =1.0, \, \epsilon_1=1.0, \, \epsilon_2=0.5,\,  \epsilon_3 =0.1,\,  g_1=g_3=g_4=0.1,\,  g_2=g_5=g_6 =0$.  (b)  Transition probabilities from one state to its time reversal counterpart state as functions of time. For this case, numerical simulations of the evolution are from time $t=-6.0$ to $t=6.0$.  Parameters:  $h_1 =0 .5, \epsilon_1=1.0, \, \epsilon_2=0.5, \,\epsilon_3 =0.1, \,  g_1=g_3=1.0, \, g_4=0.5, \, g_2=g_5=g_6 =0$. (c) Transition probabilities with initial level $3$ as function of parameter $g$ defined as $ g_1=g_3=g_4=4.0 g$. Other parameters: $g_2=g_5=g_6=0,\,  h_1 = 2.0, \epsilon_1=2, \, \epsilon_2=1, \, \epsilon_3=0$.   (d) Transition probabilities with initial level $4$  as functions of parameter $\epsilon$ defined as $\epsilon_1=4.0 \epsilon, \epsilon_2= 2\epsilon,  \epsilon_3=0$.  Other parameters: $ g_1=g_3=g_4=1.2, \, g_2=g_5=g_6=0,\,  h_1 = 2.0$.  For (c-d), numerical solution is obtained for time interval $(-T,T)$, where  $T=200$. Solid lines in (c-d) are the guides for eyes.}
\end{figure}

\subsection{ Central spin model}
 Consider now a simple example of a central spin model, in which  the central spin-$1/2$, $\hat{\bs}$, interacts with two nuclear spins-1/2, $\hat{\bs}^{(1)}, \hat{\bs}^{(2)}$. We disregard the g-factor of nuclear spins and assume that only the central spin interacts with the linearly time-dependent magnetic field. Consider the Hamiltonian that does not conserve the total spin (which happens for dipole coupling):
\be
\hat{H}_3 &=& h_1 t \hat{s}_z + \epsilon_1 \hat{s}_z \hat{s}^{(1)}_{z} + \epsilon_2 \hat{s}_z \hat{s}^{(2)}_{z} + \epsilon_3  \hat{s}^{(1)}_{z}  \hat{s}^{(2)}_{z} + g_1 \hat{s}_x \hat{s}^{(1)}_{z}  + g_2 \hat{s}_x \hat{s}^{(2)}_{z}  \nn \\
&&   + g_3 \hat{s}_x \hat{s}^{(1)}_x + g_4 \hat{s}_x \hat{s}^{(2)}_x + g_5 \hat{s}_y \hat{s}^{(1)}_y + g_6 \hat{s}_y \hat{s}^{(2)}_y. 
 \label{H3}
\ee

In the basis, $(1, 2, 3, 4, 5, 6, 7, 8) \equiv ( | \ua \ua \ua \ra, |\ua \ua \da \ra, | \ua \da \ua \ra, | \ua \da \da \ra,  | \da \ua \ua \ra, |\da \ua \da \ra, | \da \da \ua \ra, | \da \da \da \ra)$, of the eigenvectors of $\hat{s}_z \hat{s}_z^{(1)} \hat{s}_z^{(2)}$, the Hamiltonian is represented by an 8$\times$8 matrix:
\be
\hat{H}_3 \begin{pmatrix}
  \beta t + \Delta_1   & 0 & 0 & 0 & \gamma_1 & \gamma_2 & \gamma_3& 0 \\
   0 & \beta t + \Delta_2   & 0 & 0 & \gamma_4 & \gamma_5 & 0 & \gamma_3  \\
  0 & 0  & \beta t + \Delta_3 & 0 & \gamma_6 & 0 & -\gamma_5 & \gamma_2 \\  
  0 & 0  &   0 & \beta t + \Delta_4 & 0 & \gamma_6 & \gamma_4 &  -\gamma_1 \\  
 \gamma_1 & \gamma_4 & \gamma_6  & 0  & -\beta t + \Delta_4 & 0 & 0 & 0 \\
  \gamma_2 & \gamma_5 & 0              & \gamma_6  & 0& -\beta t + \Delta_3 & 0 & 0 \\
 \gamma_3  & 0               & -\gamma_5 & \gamma_4  & 0 & 0 & -\beta t + \Delta_2  & 0 \\
  0             &  \gamma_3 & \gamma_2 &  -\gamma_1  & 0  & 0 & 0 & -\beta t + \Delta_1
\end{pmatrix} ,  
\ee
where  $\beta= h_1/2$, 
$\Delta_{1,2} = \frac{\epsilon_1}{4} \pm \frac{\epsilon_2}{4} \pm \frac{\epsilon_3}{4}, \Delta_{3, 4} = - \frac{\epsilon_1}{4} \pm \frac{\epsilon_2}{4} \mp \frac{\epsilon_3}{4} $, 
$\gamma_{1, 5} = \frac{g_1}{4} \pm \frac{g_2}{4}, \gamma_{2, 4} = \frac{g_4}{4} \mp \frac{g_6}{4}, ~~\gamma_{3, 6} = \frac{g_3}{4} \mp \frac{g_5}{4} 
$.

Figure~\ref{fig2} shows that this model describes crossings between two bands of energy levels. 
This model does not satisfy the criteria of integrability discussed in  \cite{sinitsyn2015solvable,sinitsyn2015exact} because it allows interference of different trajectories with nonzero effect of the dynamic phase, as illustrated in Fig.~\ref{fig2}(a) for the transition from level-$1$ to level-$2$. Red and Black dashed lines highlight two possible semiclassical trajectories connecting these states. Such trajectories acquire different dynamic phases which influence final transition probabilities. The rest of Fig.~\ref{fig2} is analogous to Fig.~\ref{fig1}. In Fig.~\ref{fig2}(b), we
show that transition probabilities from levels  $n=1, 2, 3, 4$ to their respective time-reversed counterparts $n'=8, 7, 6, 5$ are generally nonzero but become zero by the end of the evolution, in agreement with the No-Scattering Theorem.  In Fig.~\ref{fig2}(c),  transition probabilities are shown as functions of couplings. Numerical simulations confirm that $P_{3\rar 6}  = 0$ and $P_{4\rar 5}  = 0$ (time reversal counterparts of states $3$ and $4$ are states, respectively, $6$ and $5$).  Figure~\ref{fig2}(d) shows that transition probabilities from the level-4 to other levels generally depend on the energy level splitting parameters $\epsilon_i$. In semiclassical independent crossing approximation, all transitions from level-4 do not depend on $\epsilon_i$, hence, dependence of some of the transition probabilities on those parameters indicate the breakdown of this approximation.  
 Nevertheless, the transition probability from level-4 to level-5 is identically zero in this regime in agreement with the No-Scattering Theorem.

\section{Discussion}

We formulated and proved the No-Scattering Theorem stating that  scattering amplitudes between Kramers doublet states of half-integer spin systems with {\it explicitly  time-dependent couplings that split Kramers degeneracy} are identically zero for some broad range of driving protocols. An important class of applications of this theorem is found in spin systems driven by a  linearly time-dependent magnetic field. We demonstrated on specific examples that this theorem is valid even in cases when models are not generally solvable. This is a new ``no-go" type of exact results \cite{sinitsyn2004counterintuitive} in the multistate LZ theory. We would like also to point that our result applied to a single spin-1/2 describes a special case of zero area pulses that have been used for control of quantum two-level systems \cite{Lehto2016}. 

Apart from direct physical applications, we expect that this theorem will shed  light on the currently mysterious integrability in the multistate LZ theory. It was discovered recently that  in multistate LZ models that satisfy certain conditions scattering probabilities between all pairs of diabatic states can be found analytically and exactly \cite{sinitsyn2015exact,sinitsyn2015solvable, sinitsyn-16QED}. 
Some of these models are quite complex, e.g., Ref.~\cite{sinitsyn-16QED} described solution of the time-dependent version of the Tavis-Cummings model, which time-independent counterpart is solvable only in the sense of applicability of the algebraic Bethe ansatz. 

Currently, there is no proof for the validity of LZ integrability conditions apart from numerous numerical tests for the models that have been solved by using them. However, there is an important hint that can lead to the proof: integrability conditions require the presence of a sufficient number of exact eigenenergy level crossings in the plot of the spectrum of the Hamiltonian as a function of time.
 In case of the Hamiltonian (\ref{eq:general_H}) with a half-integer total spin and conditions (\ref{eq:coeff}), existence of exact crossing points is guaranteed at $t=0$ by the Kramers theorem. Since the symmetry leading to these crossing points is well understood, we were able to identify the effect of this symmetry on the scattering matrix. We found that each crossing point corresponds to a specific constraint that, in this case, forbids transitions between some  states. 

Based on this observation, we can speculate that exact crossing points play the role in the multistate LZ theory  that is similar to the role of integrals of motion in classical mechanics. 
Each exact crossing point corresponds to some symmetry (which may be hard to identify) that constraints the scattering matrix. When the number of exact level crossings is sufficiently large, the scattering matrix becomes so constrained that the multistate LZ model becomes fully solvable. Recently, it became possible to generate families of time-dependent Hamiltonians that guarantee the presence of exact crossing points of nontrivial origin \cite{com-partner,armen}.  It should be insightful to test our conjecture in these models. 



\section{Appendix: Proof of $\Theta \hat{S}_{\alpha} \Theta^{-1} = - \hat{S}_{\alpha}$ }

Operator $e^{-i\pi \hat{S}_y}$ is the spin rotation operator around y-axis by angle $\pi$. Hence, the following identities hold:
\be
e^{-i\pi \hat{S}_y} \hat{S}_x = - \hat{S}_x e^{-i\pi \hat{S}_y} {~~ \rm and~~}  e^{-i\pi \hat{S}_y} \hat{S}_z = - \hat{S}_z e^{-i\pi \hat{S}_y},
 \label{eq:sxsz}
\ee
Also, naturally,
\be
e^{-i\pi \hat{S}_y} \hat{S}_y = \hat{S}_y e^{-i\pi \hat{S}_y}. 
\label{eq:sy}
\ee
Then we recall that  the representations of $\hat{S}_x$ and $\hat{S}_z$ are real matrices while $\hat{S}_y$-matrix has imaginary entries. Therefore, $\hat{K} \hat{S}_{x, z} = \hat{S}_{x, z}  \hat{K}$, while $\hat{K} \hat{S}_y = - \hat{S}_y \hat{K}$, and $\hat{K}e^{-i\pi \hat{S}_y}=e^{-i\pi \hat{S}_y} \hat{K}$. So 
\be
e^{-i\pi \hat{S}_y} \hat{K} \hat{S}_{\alpha} = - \hat{S}_{\alpha} e^{-i\pi \hat{S}_y} \hat{K}, \quad \alpha=x,y,z,
\ee 
which is just $\Theta \hat{S}_{\alpha} \Theta^{-1} = -\hat{S}_{\alpha}$.

\section*{Acknowledgements}
The work
was carried out under the auspices of the National Nuclear
Security Administration of the U.S. Department of Energy at Los
Alamos National Laboratory under Contract No. DE-AC52-06NA25396. Authors also thank the support from the LDRD program at LANL.


%

\end{document}